\documentclass[pre,floatfix,amssymb,twocolumn,amsmath]{revtex4-2}
\usepackage{hyperref}
\usepackage{color}
\usepackage{graphicx}
\usepackage{xspace}

\usepackage{ifxetex}
\ifxetex
\usepackage{fontspec}
\else
\usepackage[T1]{fontenc}
\usepackage[utf8]{inputenc}
\fi
\usepackage{textcomp}

\usepackage[caption=false]{subfig}

\usepackage{pgffor}
\newcommand\subfig[2]{{Fig.~\ref{#1}{#2}}}

\newcommand\subcap[1]{{(#1):}}

\newcommand{\SET}[1]{\{#1\}}

\newcommand{\eq}[1]{eq.~\eqref{#1}}

\newcommand{\Eq}[1]{Eq.~\eqref{#1}}
\newcommand{\fig}[1]{Fig.~\ref{#1}}

\newcommand{\sect}[1]{Section~\ref{#1}}

\newcommand{\etc}{\textrm{etc.}}

\newcommand{\ie}{\textrm{i.e.}}

\newcommand{\OCAL}{\mathcal{O}}  
\newcommand{\xbar}{\overline{x}}  
\newcommand{\expb}[1]{\exp \glb #1 \grb} 
\newcommand{\sinb}[2][]{\sin^{#1} \glb #2 \grb}  

\newcommand{\loga}[2][]{\log^{#1}\! \gla #2 \gra}  

\newcommand{\gla}{\,}  
\newcommand{\gra}{}  
\newcommand{\glb}{\left(}  
\newcommand{\grb}{\right)}  

\newcommand{\TO}{,\ldots,}
\newcommand{\VEC}[1]{\mathbf{#1}}
\newcommand{\ehatvec}{\hat{\VEC{e}}}

\newcommand{\qvec}{\VEC{q}}

\newcommand{\vvec}{\VEC{v}}

\newcommand{\xvec}{\VEC{x}}

\newcommand{\mean}[1]{\left\langle #1 \right\rangle}
\newcommand{\half}{\frac{1}{2}}

\newcommand\bigOb[1]{\ensuremath{\OCAL\glb #1 \grb}}

\newcommand\bigObs[1]{\ensuremath{\OCAL  (#1)}}

\newcommand{\PARTICLE}{sphere\xspace}
\newcommand{\PARTICLES}{spheres\xspace}

\newcommand{\oned}{1D\xspace}
\newcommand{\twod}{2D\xspace}
\newcommand{\xdir}{\ehatvec_x}
\newcommand{\ydir}{\ehatvec_y}
\newcommand{\tchain}{\tau_{\text{chain}}}
\newcommand{\lchain}{\Delta_{\text{chain}}}
\newcommand{\hff}{h_{\text{{f}{f}}}}
\newcommand{\Vff}{V_{\text{{f}{f}}}}
\newcommand{\wlane}{w_{\text{lane}}}
\newcommand{\Struct}[1]{S_{#1}}
\newcommand{\hardsphere}{\text{hs}}

\begin{document}
\title{Large-scale dynamics of event-chain Monte Carlo}

\author{A.~C.~Maggs} \email{anthony.maggs@espci.fr} \affiliation{CNRS UMR7083,
  ESPCI Paris,
  Université PSL, 10 rue Vauquelin, 75005 Paris, France}

\author{Werner Krauth} \email{werner.krauth@ens.fr} \affiliation{Laboratoire de 
Physique de l’Ecole normale supérieure, ENS, Université PSL, CNRS, Sorbonne 
Université, Université de Paris, Paris, France}

\begin{abstract}
Event-chain Monte Carlo (ECMC) accelerates the sampling of hard-sphere systems, 
and has been generalized to the potentials used in classical molecular 
simulation. Rather than imposing detailed balance on transition probabilities, 
the method enforces a weaker global-balance condition in order to guarantee 
convergence to equilibrium. In this paper we generalize the factor-field variant 
of ECMC to higher space dimensions. In the two-dimensional fluid phase, 
factor-field ECMC  saturates the lower bound $z=0$ for the dynamical scaling 
exponent for local dynamics, whereas molecular dynamics is characterized by 
$z=1$ and local Metropolis Monte Carlo by $z=2$. In the presence of hexatic 
order, factor fields are not found to speed up the convergence. We indicate
applications to the physics of glasses, and note that generalizations of factor 
fields could couple to orientational order.
\end{abstract}
\maketitle

\section{Introduction}
\label{sec:Introduction}

Event-chain Monte Carlo (ECMC) has led to important advances in the simulation  of $N$-particle 
systems~\cite{Bernard2009,Peters_2012,Michel2014JCP, Kampmann2015, 
Harland2017,Hu2018, Klement2019,Michel2020, Krauth2021eventchain,Kampmann2021}. 
The efficiency gains that it brings have led to an improved  understanding of phase transitions in two 
spatial dimensions (2D)~\cite{Bernard2011}. As a  non-reversible Markov-chain 
Monte Carlo (MCMC) algorithm~\cite{SuwaTodoPRL2010, Turitsyn2011, 
BierkensPDMC2017}, ECMC exactly samples the equilibrium Boltzmann distribution. 
However, it is itself out of equilibrium, because it replaces the diffusive 
dynamics of reversible MCMC (based on the detailed-balance condition) by 
ballistic dynamics (rooted in the more general global-balance condition). 
Non-reversible MCMC can approach the steady state, often the equilibrium 
Boltzmann distribution, on shorter time scales than reversible 
formulations~\cite{Diaconis2000,Chen1999}. At large MCMC times, steady-state 
autocorrelation functions are exponential both for reversible and generally also 
for non-reversible Markov chains. The slowest mode generally
relaxes on a time scale $\tau$ which depends 
on the system size $L$ as $\tau \sim L^z$. For $N$-particle systems in one 
spatial dimension (\oned), ECMC can be analyzed in great 
detail~\cite{KapferKrauth2017,Lei2019} and compared to molecular dynamics and to 
the reversible local Metropolis algorithm. The autocorrelation functions of 
density fluctuations in ECMC, as in molecular dynamics,  are characterized by a 
dynamic exponent $z=1$, where the unit of time corresponds to a sweep of $N$ 
moves or events. This is asymptotically faster than for the reversible local 
Metropolis algorithm,  for which $z=2$ so that the autocorrelation time, in $d$ 
dimensions, corresponds to $\sim L^z N \sim N^{1 + z/d}$ moves. For 
\oned systems, a powerful variant of ECMC~\cite{Lei2019} adds a factor potential 
to the Hamiltonian. The factor potential  leaves thermodynamic properties 
rigorously invariant, yet takes the system to zero pressure, $P$ by replacing 
the external forces by an attraction between particles. Factor-field  
ECMC lowers the \oned dynamic exponents to  $z=1/2$, the theoretical 
minimum for a local MCMC algorithm. This acceleration is accompanied by 
super-diffusive dynamics of the instantaneous active 
particle~\cite{Lei2019,KimuraHiguchi2017}.

In the present paper, we formulate factor fields for higher-dimensional particle 
models and implement them for hard spheres in a \twod box. In fluid phases, 
hydrodynamic fluctuations that are coupled to local conservation laws constitute 
the long-lived modes for stochastic dynamics of the types realized in local MCMC 
algorithms~\cite{Martin_1972,Forster1975,HohenbergHalperin1977}. We demonstrate 
through extensive numerical simulations for the implemented \twod hard-sphere 
model that factor fields can again lower the dynamical scaling 
exponents for such 
modes to their theoretical minimum, below those reached by molecular 
dynamics and by reversible local Monte Carlo. The reduction in dynamical scaling 
exponents translates into shorter correlation times for density fluctuations 
and, more generally, shorter overall correlation times. 
 
The \twod factor fields introduced in this paper do not seem to couple to 
orientational degrees of freedom. In the hexatic phase, orientational order is 
itself (quasi)-long-ranged, and the dynamical scaling exponent of the 
hexatic field are thought to be
diffusive for Hamiltonian dynamics, with $z \sim 2$ 
(see~\cite{Zippelius1980}). We expect this scaling to hold for reversible MCMC 
and for ECMC, but also for molecular dynamics.
The computation of  dynamical scaling exponents in the ordered phases beyond 
initial explorations~\cite{Weigel2018} remains an unsolved problem, which is 
much more difficult than establishing the phase diagrams. 
Devising ECMC with modified factor fields with 
reduced scaling exponents for ordered phases  appears as an outstanding 
challenge. 
We point to 
possible applications of the present formalism, for example in the study of 
glasses, where current simulation methods~\cite{Ninarello2017} use a non-local 
MCMC algorithm to relax structural degrees of freedom, but molecular dynamics to 
relax density modes. ECMC with factor fields may well decrease dynamical scaling 
exponents in these higher-dimensional polydisperse-particle models, and thus 
open the way for simulations of larger glassy systems than currently possible. 

The ECMC algorithm evolves in  continuous MCMC time $t$. Its event-driven 
implementation eliminates all discretization 
errors~\cite{Bernard2009,Peters_2012,Michel2014JCP}. In the straight variant of 
ECMC, a unique active particle moves at unit speed in the chain direction, along 
one of the coordinate axes, in \twod between $\xdir$ and $\ydir$. At a lifting 
event, the motion transfers from the active sphere to the  target sphere, 
preserving the chain direction. For hard spheres, such lifting events correspond 
to pair collisions. The algorithm is organized into event chains of chain time 
$\tchain$,  an intrinsic parameter of straight ECMC that influences its 
efficiency. At the end of a chain a new active particle is randomly sampled and 
the chain direction may alternate between $\xdir$ and $\ydir$. The active sphere 
changes identity at each lifting move. It advances at an average speed which is 
proportional to the pressure, $P$. More precisely, the total displacement 
$\lchain$ of the chain---the difference of the final position of the last chain 
sphere and of the initial position of the initial chain sphere---depends on the 
continuous MCMC time $\tchain$ of the chain as~\cite{Michel2014JCP}
\begin{equation}
\beta P =   \frac{N}{V}\ \mean{\frac{\lchain }{ \tchain}},
\label{equ:FabulousFormula}
\end{equation}
where $\mean{.}$ is the ensemble mean and $\beta$  the inverse temperature, 
which is set to $\beta=1$ throughout. \Eq{equ:FabulousFormula} holds for general 
pair potentials and it allows for the presence of factor 
potentials. 

In this paper, we focus on the two-dimensional system of $N$ hard spheres of 
radius $\sigma$ in a square box of sides $L$ with periodic boundary condition. 
The position of each sphere is given by the coordinates of its center. The 
density is $\eta = N \pi \sigma^2 / L^2$. For large $N$, the system is fluid for 
densities $0<\eta<0.7$. It is in fluid--hexatic coexistence for $0.7 <\eta< 
0.716$ as a consequence of an underlying first-order phase 
transition, and it is hexatic at $0.716 < \eta \lesssim 0.72$, above which it is 
solid~\cite{Bernard2009, ThorneyworkDullens}. The hard-sphere factor-field 
algorithm can be  generalized to smooth potentials, where we expect our 
conclusions to carry over.

\section{Factor fields in \oned and \twod}
\label{sec:FFonedtwo}

We first consider  $N$ spheres on a continuous \oned interval of length $L$. The 
hard-sphere pair interaction,
\begin{equation}
V_\hardsphere = 
\sum_{i=1}^N v_{\hardsphere}( x_{i+1}-x_{i}), 
\end{equation}
between successive spheres (with $v_{\hardsphere}$ either zero or infinity) is 
understood with periodic boundary conditions in positions ($x + L \equiv x$) and 
indices ($i+ N \equiv i$). The \oned factor potential~\cite{Lei2019} consists in 
a sum of linear potentials 
\begin{equation}
  \Vff = -\hff \sum_i (x_{i+1} -x_i ) = -\hff L, 
\label{equ:ff_definition} 
\end{equation}
that is constant for any factor field $\hff$,   because of the periodic 
boundary conditions. The factor potential $\Vff$ can be added to the 
inter-particle potential without changing correlation functions, as the constant, \(-\hff L\),
cancels between the statistical weight and the partition function. 
Furthermore, force-based time evolutions such as molecular dynamics and 
energy-based Monte Carlo trajectories (Metropolis, heatbath, \etc) have 
indistinguishable dynamics for all $\hff$. In contrast, in ECMC, the acceptance 
of a move depends on independent decisions made by pairs of spheres 
(see~\cite{Krauth2021eventchain} for a detailed discussion of the general case), 
and the ECMC dynamics is strongly altered through $\Vff$. A factor field $\hff = 
P$, with $P$ throughout this  paper the pressure in the 
absence of factor fields, implies that the factor-field system, with potential 
$V_\hardsphere + \Vff$, has zero pressure, so that the  average chain 
displacement $\mean{\lchain}$ vanishes (see \eq{equ:FabulousFormula}). The 
lifting move between an active and a target sphere can be a hard-sphere 
collision, that always goes forward in the chain direction, or else a 
factor-field lifting move that always goes backward. In the absence of drift, at 
$\hff = P$, the position of the instantaneous active sphere (that changes 
identity at each lifting move)  is characterized by hyperdiffusive motion with 
long-term memory. In the steady state, this lowers the \oned dynamical exponent 
from $z=1$ to $z=\half$ and it also accelerates mixing. 

\begin{figure}
  \includegraphics[width= \columnwidth]{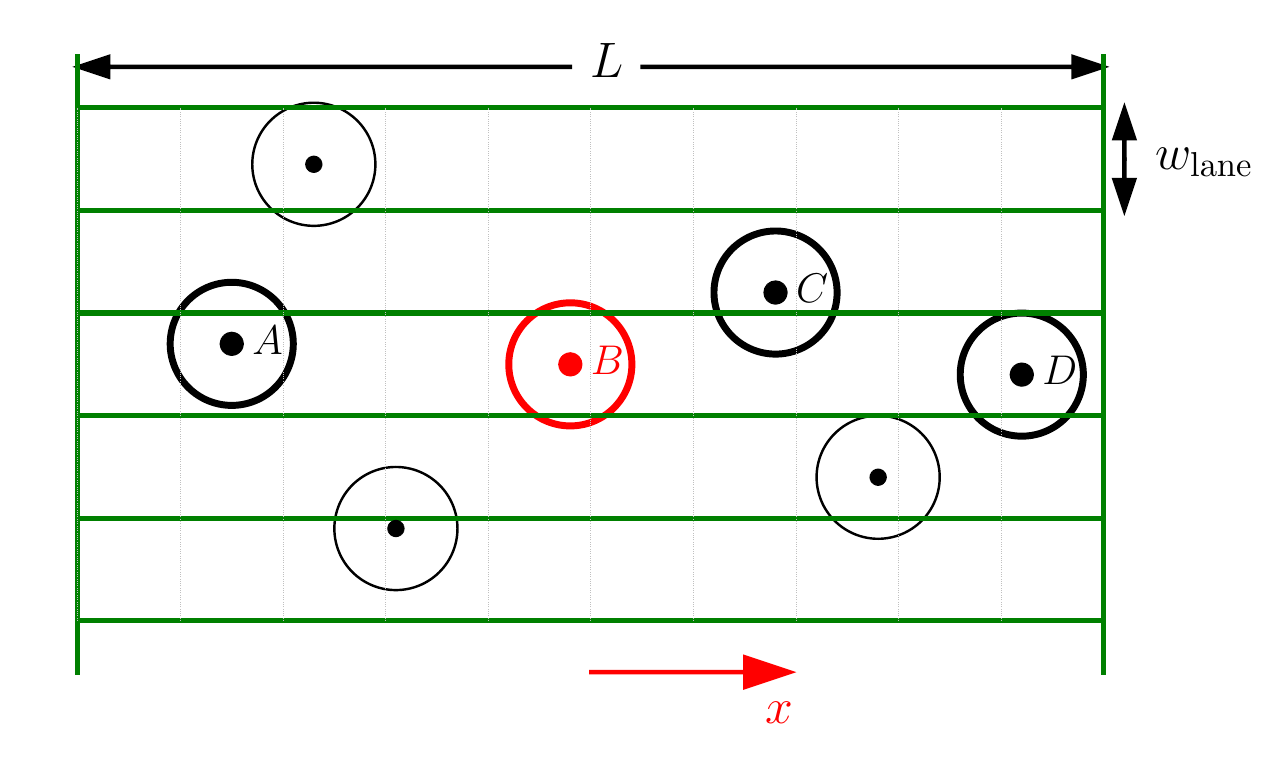}
  \caption{Spheres in a \twod box,  with a 
grating of lanes 
  derived from a \twod cell system.
   Hard-sphere lifting moves  correspond to collisions (for 
example of the active sphere $B$ with 
   $C$). Factor-field lifting moves always concern 
same-lane spheres (for example $A$, for the active sphere $B$).}
\label{fig:active}
\end{figure}

To adapt factor fields to \twod hard spheres, we construct for each chain a 
grating of lanes of width $\wlane \lesssim 2\sigma$ that is compatible with $L$  and 
that is oriented parallel to the chain direction (see \fig{fig:active}). 
 There,  spheres are 
quasi-one-dimensional, and the factor potential of \eq{equ:ff_definition} can 
again be added between spheres in the same lane. Spheres in nearby lanes only interact through the hard-sphere 
potential (see \fig{fig:active} for examples). The factor potential is now a sum 
over all lanes of an expression analogous to the \oned factor potential of 
\eq{equ:ff_definition}, each of which is a constant.  Again, the factor field 
leaves thermodynamic properties rigorously invariant. As we will show, at least 
in the fluid phase, it also lowers the dynamical scaling exponent to its 
theoretical minimum. 

\begin{figure}
  \includegraphics[width=0.8\columnwidth]{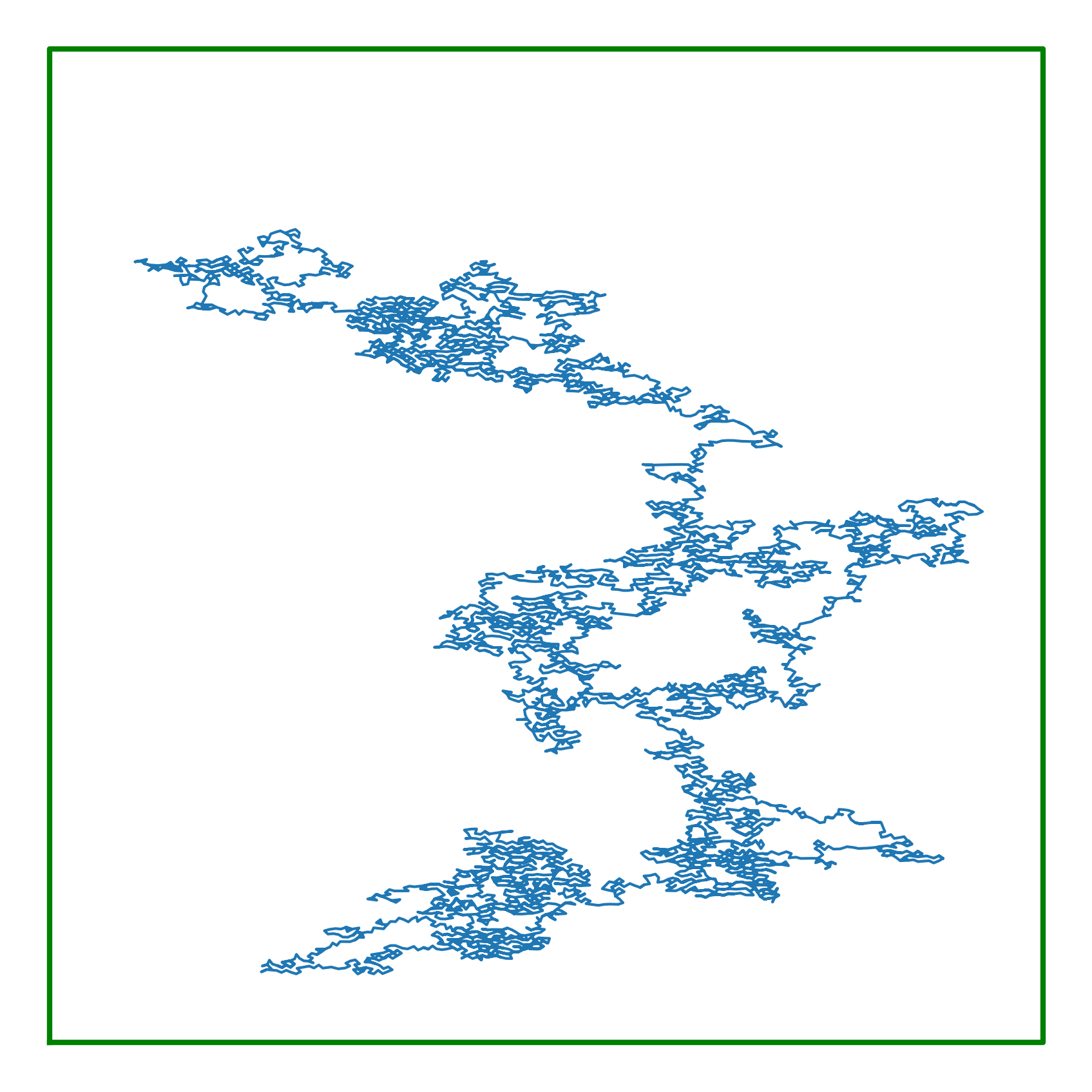} 
  \caption{Active-sphere trajectory featuring anisotropic 2D diffusion of a 
single event chain with direction $\xdir$ for the \twod hard-sphere system at 
$\eta=0.67$. The optimal $\hff   = P \wlane$ is used ($N=192^2$).
  }\label{fig:trajectory}
\end{figure}

A hard-sphere lifting move can concern an active and a target sphere in 
different lanes so that the active sphere effectively moves in \twod. A 
factor-field lifting move, in contrast, always remains within a given lane. The 
optimal factor field is now $\hff=P \wlane$. At this value, the active sphere 
undergoes disordered \twod motion (see \fig{fig:trajectory}). The degree 
of anisotropy depends on the lane width $\wlane$.

Practically, the grating is derived from the \twod local cell system which is 
used to scan for possible hard-sphere collisions (see \fig{fig:active}). The 
underlying Poisson process for the factor-field events does not require the 
advance  knowledge of the position or identity of the target sphere (as the lifting move $B$ to $A$ in 
\fig{fig:active}). A factor-field event simply requires walking back through the 
cells of the active-sphere lane to find the target sphere. Narrow lanes 
($\wlane < 2 \sigma$) simplify the implementation of this algorithm, which 
however generalizes in a straightforward way to larger lanes, to arbitrary pair 
potentials and to more than two dimensions. In \twod, two
different chain directions, as  $\xdir$ and $\ydir$, are needed for the 
irreducibility of the ECMC algorithm (see~\cite{Hoellmer2021sparse} for a 
detailed discussion of irreducibility in \twod hard-sphere systems).

\begin{figure}
  \includegraphics[width=\columnwidth]{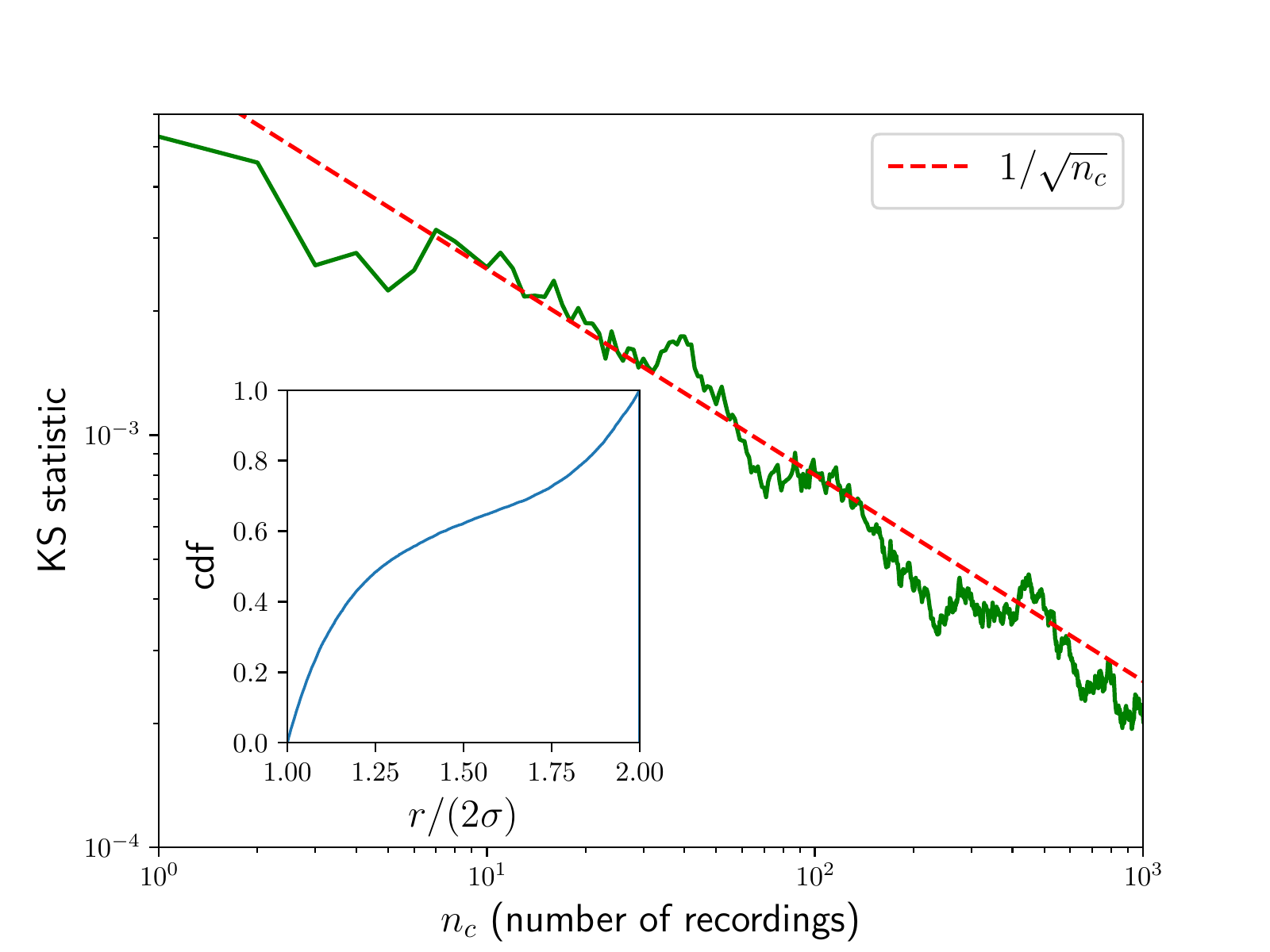}
  \caption{Maximum distance of the empirical cumulated distribution function of 
$r g(r)$ for $2\sigma < r < 4 \sigma$, with and without factor field for $n_c$ 
recordings (\ie\ $n_c$ $N$-sphere samples). The $\sim 1/\sqrt{n_c}$ scaling of 
this   Kolmogorov--Smirnov-like statistic indicates that $r g(r)$ (shown in the 
inset) is independent of $\hff$ ($N=64^2$, $\eta=0.67$). }
  \label{fig:ks}
\end{figure}

In \twod, the $P$ can be estimated through independent simulations in small physical systems. It 
need not be known to high precision to obtain efficient acceleration of the simulation.  
The correctness of the factor-field algorithm 
can be checked by comparing the pair-distribution function $ r g(r) $ near 
contact using a Kolmogorov--Smirnov-like statistic~\cite{feller}. The 
invariance of this key structural observable as a function of the factor field $\hff=0$ or 
$\hff =  P$ holds to high precision (see \fig{fig:ks}).

\section{Unidirectional displacements: Eigenmodes and dynamics}
\label{sec:PolytopeDynamics}

We now consider the restricted ECMC dynamics for a chain direction $\xdir$ and 
for moves from a specific equilibrated \twod hard-sphere configuration 
$\xvec(t=0)$. The Markov chain then evolves for $t \to \infty$ towards a 
restricted Boltzmann equilibrium among samples that can be reached from 
$\xvec(t=0)$. We write $\xvec(t) = \SET{\xvec_1(t) \TO \xvec_N(t)}$, where 
$\xvec_i(t) = \SET{x_i(t), y_i(t)}$ describes the \twod position of sphere $i$, 
with periodic boundary conditions understood, and with all $y_i$ independent of 
$t$. (See \sect{sec:FullDynamicsLiquids} for the full dynamics, with chain 
directions $\xdir$ and $\ydir$.) As discussed, spheres inside a narrow lane 
cannot re-order. The same applies to pairs $i,j$ of spheres in different lanes, 
with  $| y_i - y_j| < 2 \sigma$. Each resulting constraint between $x_i$ and 
$x_j$ can be expressed as an inequality, and the sample space accessible from 
$\xvec(t=0)$ forms a convex polytope~\cite{KapferPolytope2013}. The ECMC 
dynamics of this restricted problem will allow us to better understand the full 
dynamics of the \twod fluid.

\begin{figure*}[!htbp]
  \begin{center}
    \includegraphics[width=\textwidth]{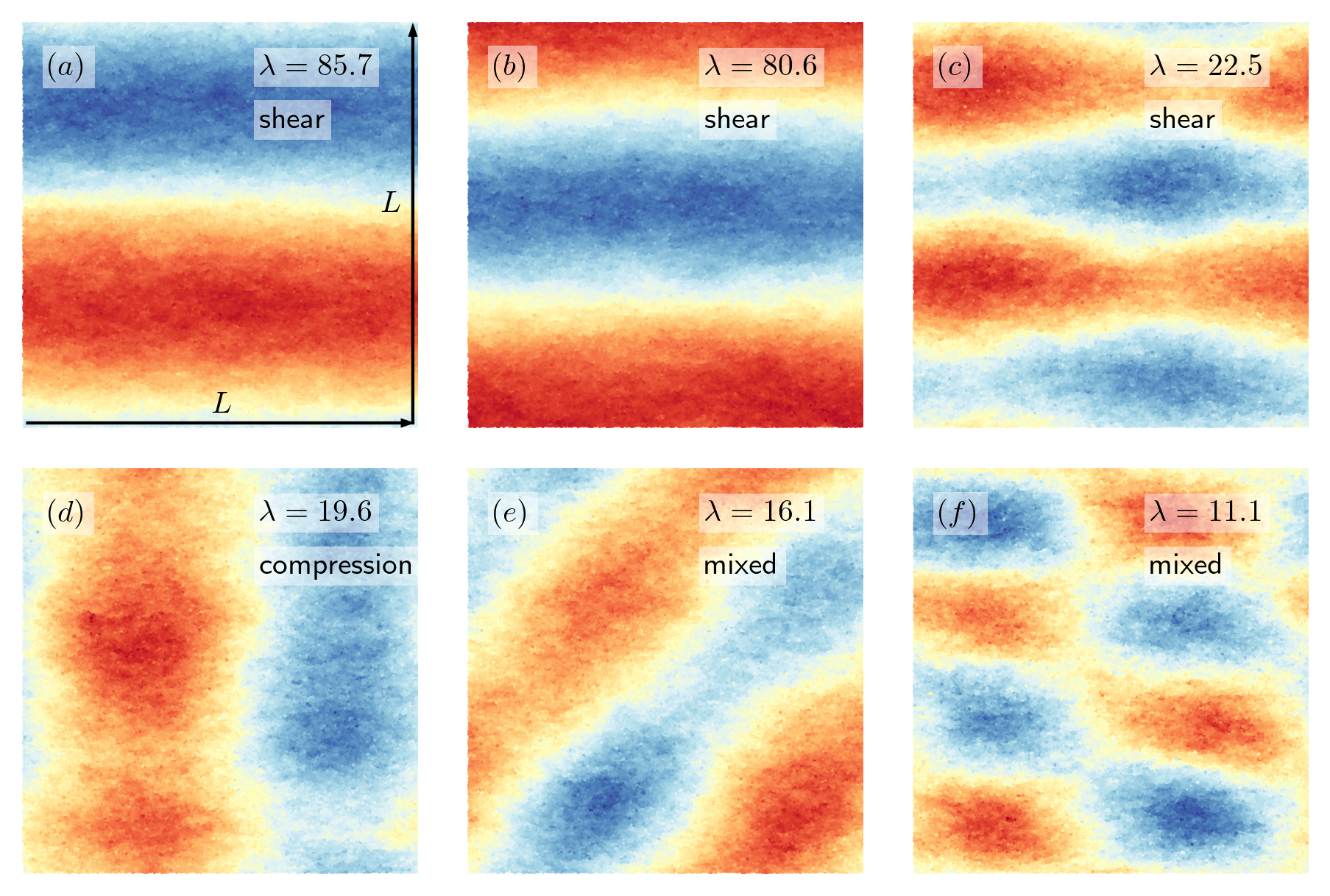}
 \end{center}
 \caption{Eigenmodes $v^{(k)}$ with dominant eigenvalues 
$\lambda^{(k)}$ of the steady-state equal-time correlation matrix of 
\eq{equ:matrix} (moves in $\xdir$ from a given initial configuration 
$\xvec(t=0)$). Red and blue denote positive and negative displacements of $x_i$ 
around its mean. Simple shears ((a), (b)) coexist with higher-order shear ((c)), 
compression ((d)) and mixed modes ($N=256^2$, $\eta=0.67$). 
\end{figure*}

We first extract the eigenmodes of  fluctuations from a given 
initial configuration $\xvec$ (moving only along $\xdir$). Subtracting the 
center-of-mass motion, allows one to define average 
positions,
\begin{equation}
    \xbar_i =\frac{1}{m} \sum_{t=1}^m x_i(t),
\end{equation}
and the equal-time correlation matrix $D= (D_{jl})$ with
\begin{equation}
  D_{jl} = \frac{1}{m} \sum_{t=1}^m (x_j(t) - \xbar_j) (x_l(t)-\xbar_l ).
  \label{equ:matrix}
\end{equation}
Lanczos' algorithm yields the largest eigenvalues $\lambda^{(k)}$ 
and eigenmodes $v^{(k)}$  of the 
symmetric $N \times N$ matrix $D$ (see 
\fig{fig:eigen} for examples). The matrix $D$, and therefore the precise 
eigenmodes, depend on the initial configuration $\xvec(t=0)$, and degeneracies 
of the associated eigenvalues, for example of the two simple shear modes, are 
lifted for this reason.

\begin{figure*}[!tbp]
    \subfloat[]{\includegraphics[width=0.33 \textwidth]{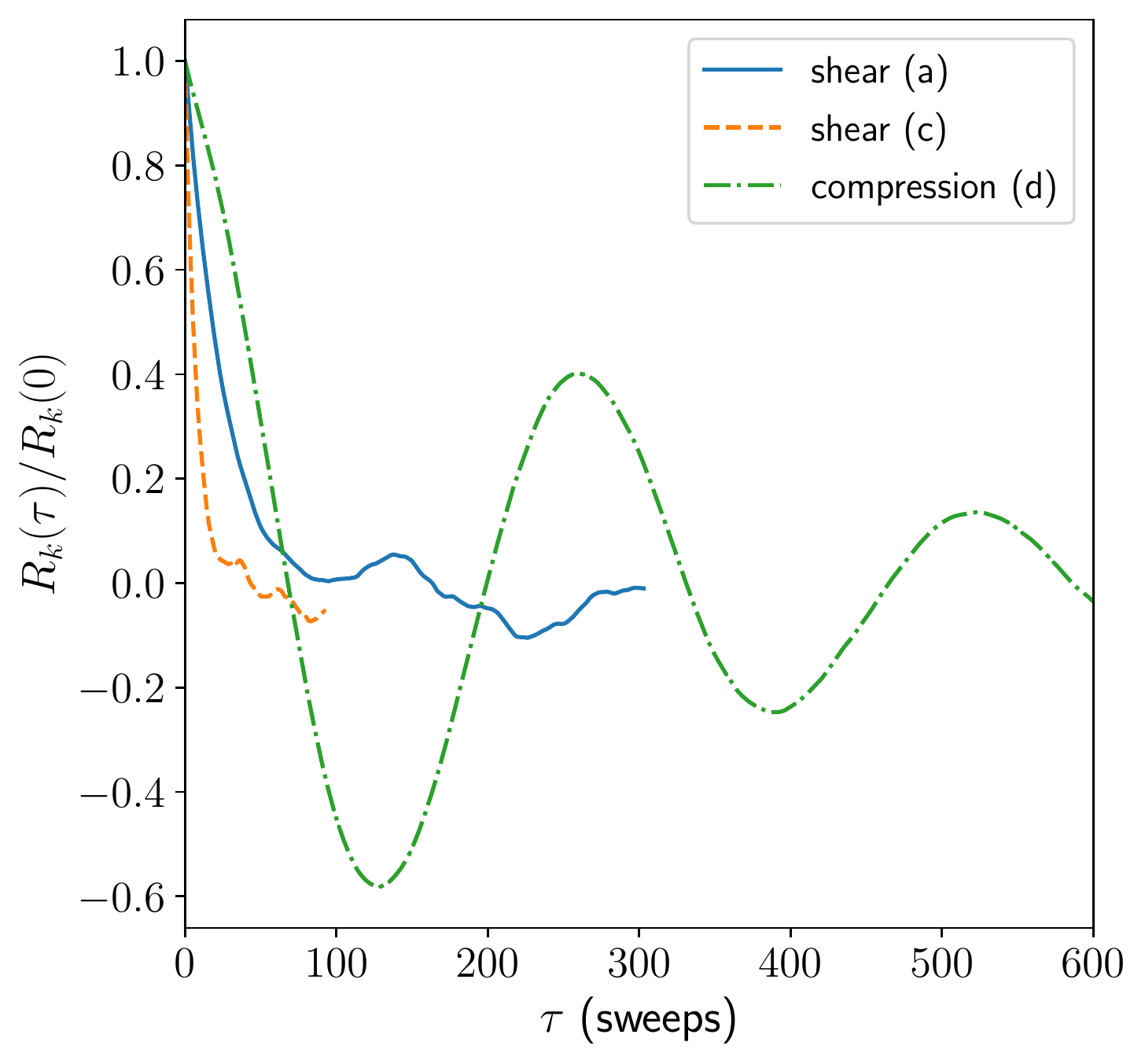}}
    \subfloat[]{\includegraphics[width=0.33 \textwidth]{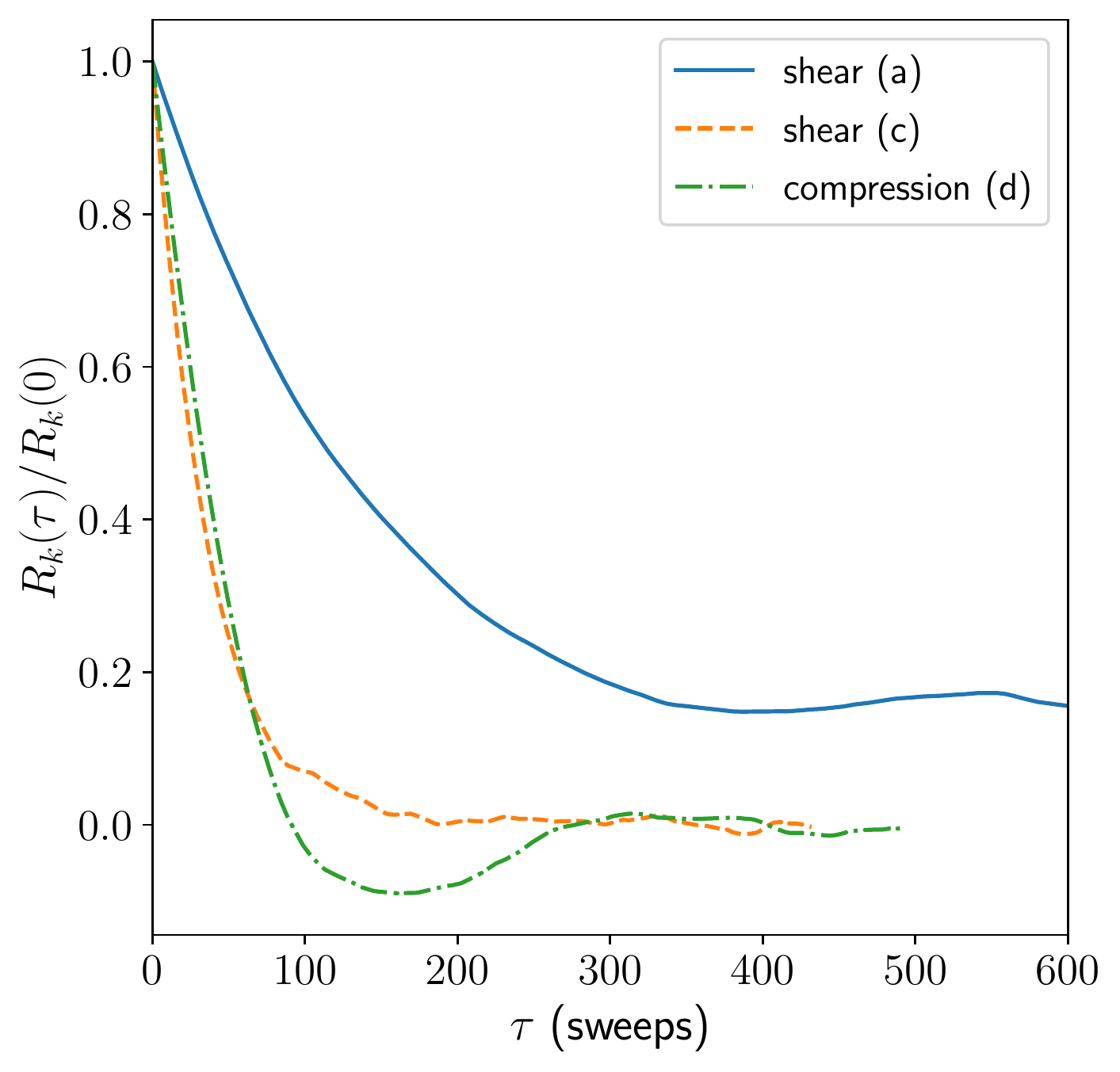}}
    \subfloat[]{\includegraphics[width=0.33 \textwidth]{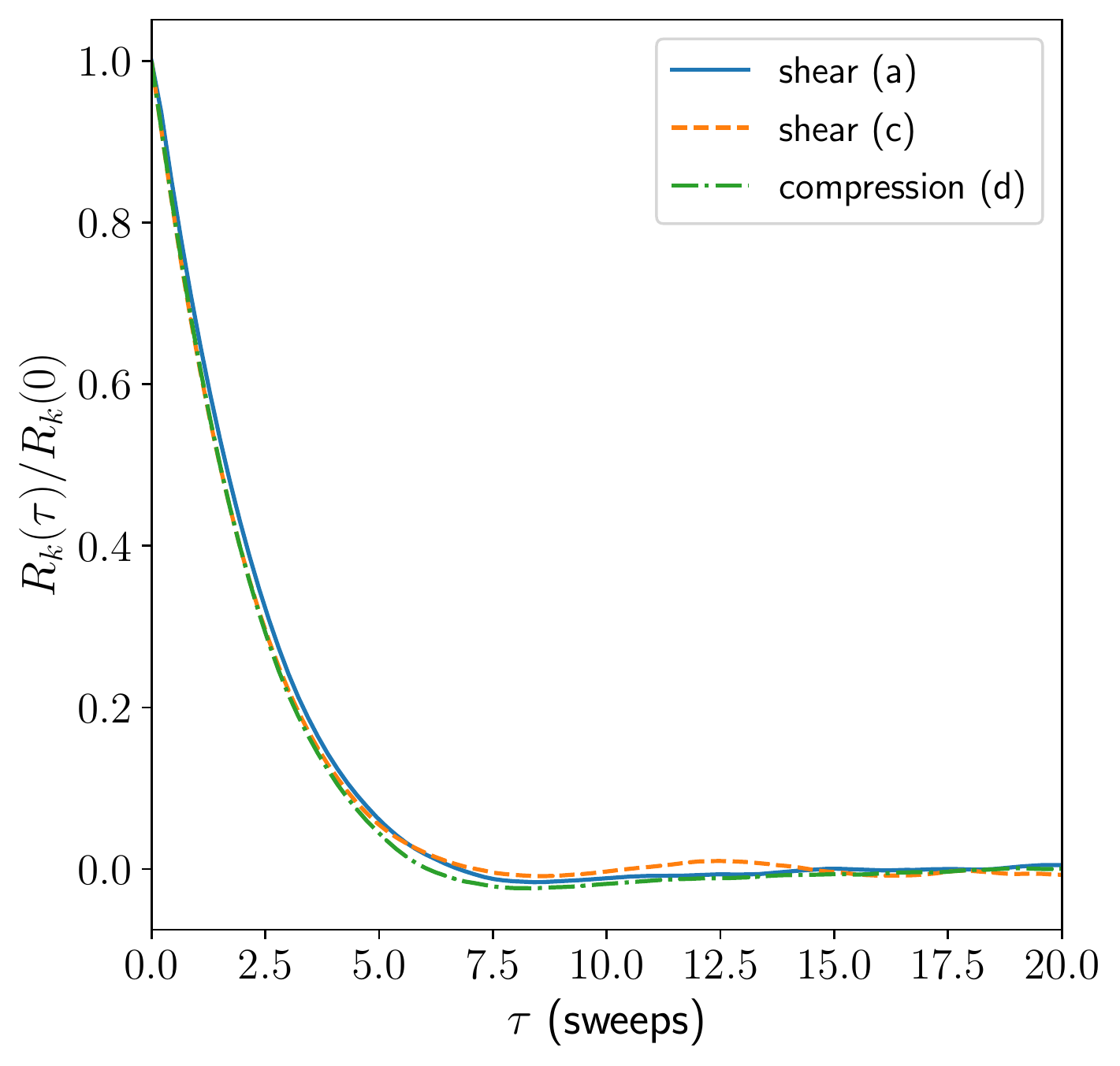}}
    \caption{Autocorrelation of the static eigenmodes 
from \eq{equ:pol} (legends 
    refer to \fig{fig:eigen}, displacements in $\ehatvec_x$ from a given 
     initial configuration $\xvec(t=0)$, center-of-mass motion subtracted).
    \subcap{a} Factor field $\hff=0$, large chain times $\tchain$ 
(corresponding to     $\lchain=3.1 L$): Compression eigenmodes relax more 
slowly that shear     eigenmodes and are oscillatory.   \subcap{b}
    Factor field $\hff=0$, small $\tchain$ (corresponding to 
$\lchain=0.54L$): 
    Shear eigenmodes may relax more slowly than compression eigenmodes.
    \subcap{c} Factor fields $\hff = P \wlane$, all
    eigenmodes decay rapidly, on similar time scales ($N=256^2$, $\eta=0.67$). 
    }\label{fig:mode}
  \end{figure*}

In a Markov chain started from the same initial configuration $\xvec(t=0)$ that was 
used to compute the correlation matrix $D$, the configurations $\xvec(t)$ (with 
all the $y_i$ kept fixed) may now  be decomposed onto the eigenmodes $v^{(k)}$. 
The time averages of autocorrelations of the eigenmode-amplitudes $a^{(k)}$ 
\begin{equation}
  R_k(\tau)= \mean{ a^{(k)}(t)\, a^{(k)}(t+ \tau) }
\label{equ:pol}
\end{equation}
characterize the decay of correlations. In straight ECMC, the chain time 
$\tchain$ is an intrinsic parameter which through \eq{equ:FabulousFormula} is 
connected to the chain length $\lchain$, its overall extension. Shortest 
autocorrelation times are obtained for a chain length $\lchain \sim \sqrt{N}\sim 
L$ (see \cite{Bernard2009}). For large chain times (chain length $\lchain \gg 
L$), compression eigenmodes relax more slowly than shear eigenmodes and show 
long-time oscillations, while for short chains ($\lchain \ll L$), the opposite 
is true, and shears can relax more slowly (see \subfig{fig:mode}{a} and 
\subfig{fig:mode}{b}). A factor field $\hff = P \wlane$ leads to the coordinated 
decay of autocorrelation functions for all eigenmodes on a time scale that is 
much shorter than for $\hff = 0$ (see \subfig{fig:mode}{c}). The sampling of the 
polytope is thus greatly accelerated by the factor fields.

\begin{figure}
    \includegraphics[width= 0.9 \columnwidth] {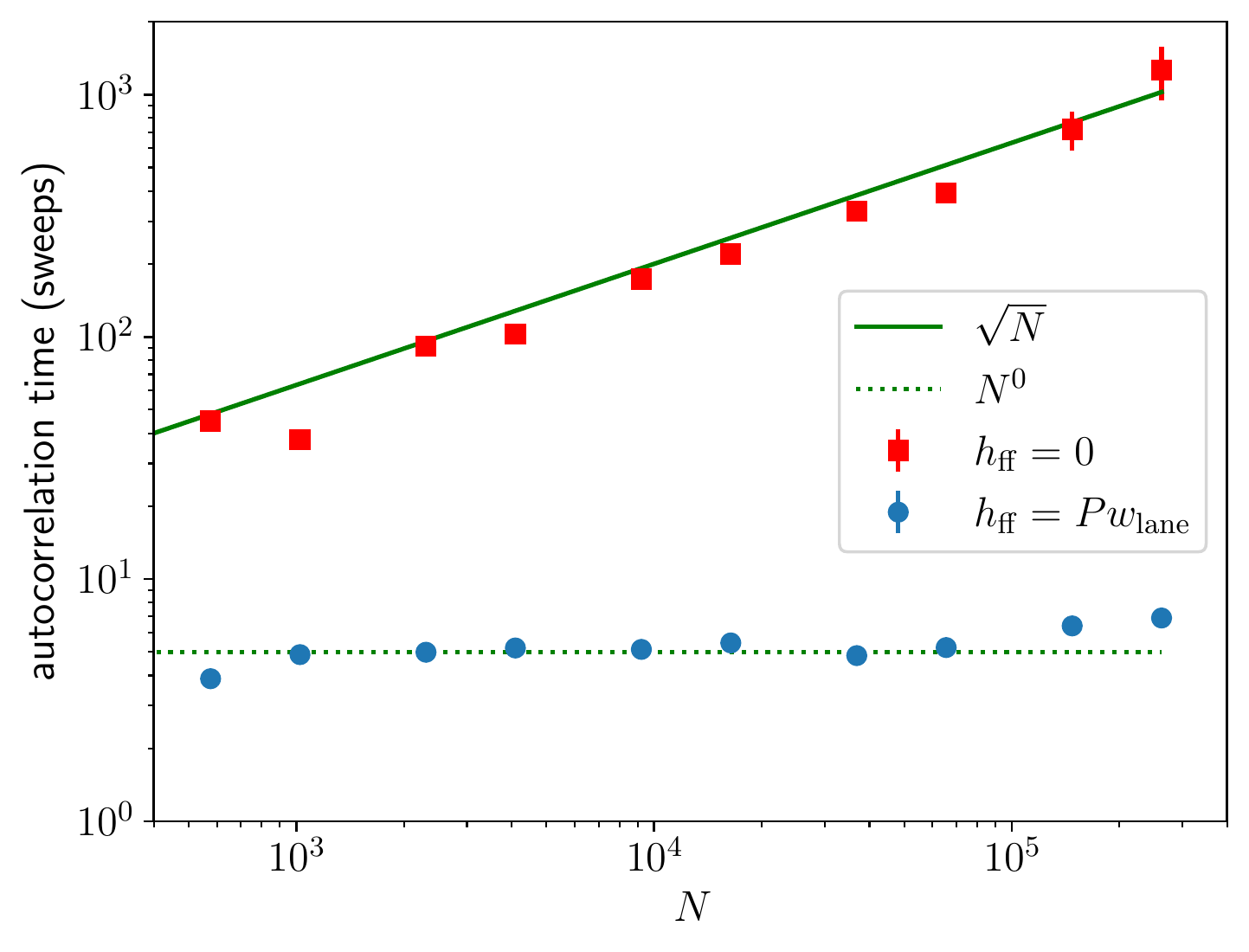}
    \caption{Integrated ECMC autocorrelation time for the eigenmodes of 
\subfig{fig:eigen}{a} and \subfig{fig:eigen}{b} in the approximation of 
\eq{equ:EigenModeFourierApprox}. (Chain direction $\xdir$, moves from  given 
initial configurations $\xvec(t=0)$.) The scaling exponent for $\hff = P \wlane$ 
appears to saturate the lower bound $z=0$ (density $\eta=0.67$).}
\label{fig:ffworks}
\end{figure}

In order to extract the integrated autocorrelation time of the dominant 
eigenmode for large  system sizes $N$ (where the equal time correlation matrix 
$D$ of \eq{equ:matrix} can not be easily stored in main memory), we model the eigenmode 
in \subfig{fig:eigen}{a}, that is the displacement $x_i = \xbar_i + v_i^{(1)}$ 
as a constant in $\xdir$ multiplied by a sine wave in $\ydir$ (see 
\subfig{fig:eigen}{b}):
\begin{equation}
\vvec^{\text{approx}} = \SET{v_1 \TO v_N} \quad \text{with}\ v_i = \sinb{2 \pi 
y_i / L}.
\label{equ:EigenModeFourierApprox}
\end{equation}
We then compute a projection coefficient as the scalar product of the 
displacement 
$\SET{(x_1(t) - \xbar_1) \TO (x_N(t) - \xbar_N)}$ with $\vvec^{\text{approx}}$. 
The timeseries of the projection coefficients yields an autocorrelation 
function, and an autocorrelation time. For $\hff  =0$, the optimal choice of the 
intrinsic parameter $\tchain$ is adopted from \subfig{fig:eigen}{b}. The 
autocorrelation time increases proportionally to $L$. For this restricted MCMC 
with fixed chain direction, this is consistent with a dynamical scaling exponent 
$z=1$ (see \fig{fig:ffworks}). For the optimal factor field $\hff=P \wlane$, the 
autocorrelation time of the approximate eigenmode of 
\eq{equ:EigenModeFourierApprox} is consistent with $z=0$. For our largest system 
with $N=512^2$, factor fields accelerate the decorrelation by more than two 
orders of magnitude. It thus appears that factor-field ECMC realizes the optimal 
$z$.

\section{ECMC Dynamics in the fluid phase}
\label{sec:FullDynamicsLiquids}

In the fluid phase, hydrodynamic modes are long-lived due to 
the presence of local conservation laws~\cite{Martin_1972,Forster1975,HohenbergHalperin1977}. Basic 
thermodynamics stipulates that fluctuations of extensive quantities, as the 
volume, grow as their square root. If a volume $V$ corresponds to a length scale 
$L^d$, then $V+ \sqrt{V}$ corresponds to a length $L + L ^{-d/2 + 1}$. In \oned, 
a test volume $\sim L$ may thus expand by its square root, leading to a lower 
limit $z = 1/2$ for local algorithms with moves on a scale \bigObs{1} This, as 
discussed, is realized by factor-field ECMC. In \twod, the test volume expands 
only by a constant length, corresponding to a minimum of a single \bigObs{1} 
move per sphere, which corresponds to $z=0$. We will now present evidence 
showing that this value of $z=0$ (which may contain a logarithm so that the 
correlation time is of order \bigObs{\log N}) is actually realized by 
factor-field ECMC, implying that a local algorithm may well reach the same 
scaling as the non-local MCMC algorithms, which have a proven mixing rate of 
\bigOb{N \loga{N}} at small but finite densities~\cite{Kannanrapidmixing2003}, 
in the fluid phase~\cite{Helmuth2020}. 

\begin{figure}[!tbp]
  \includegraphics[width=0.9 \columnwidth]{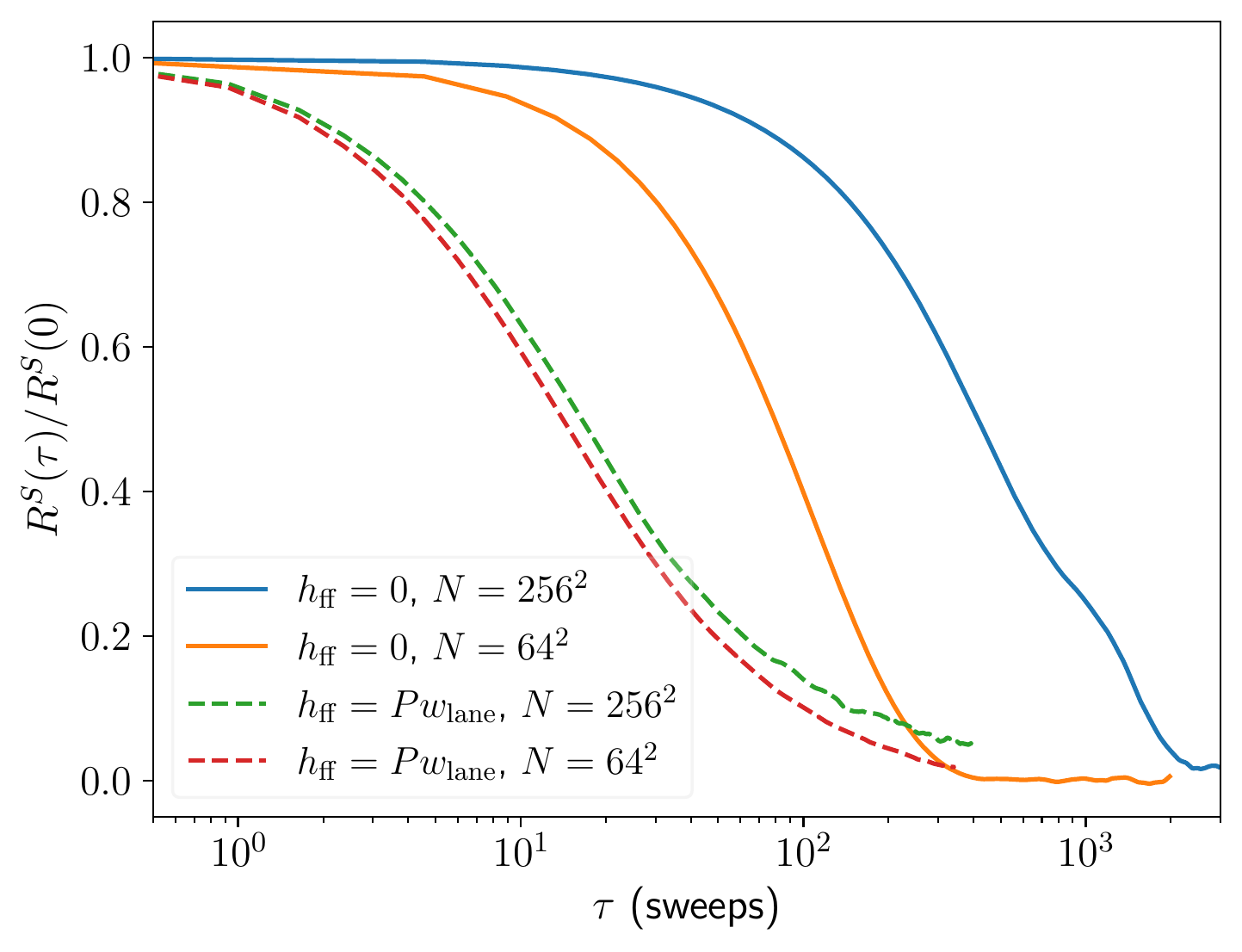}
  \caption{Autocorrelation $R^{S}(\tau)$ of the structure factor 
$\Struct{\qvec}$ for full ECMC in the fluid phase (density $\eta=0.67$). With 
factor fields, the autocorrelations seem to decay on a time scale that is 
independent of $N$, indicating $z=0$. } 
\label{fig:densf}
\end{figure}

To trace the ECMC evolution of density fluctuations at the largest available 
length scales, we consider the Fourier coefficient $\rho(\qvec)$ of the number 
density $\rho(\xvec) = \sum_i \delta(\xvec - \xvec_i)$,
\begin{equation}
  \rho( \qvec ) = \sum_j \expb{i \qvec \cdot \xvec_j},
\end{equation}
at the longest wavelength $\qvec = (2 \pi/L) (1,0)$. We studied the autocorrelation 
function of $\rho(\qvec)$ or relatedly, the autocorrelation of the corresponding 
structure factor $  \Struct{\qvec}= |\rho(\qvec)|^2 $,
\begin{equation}
  R^{S}(\tau) = \mean{ \Struct{\qvec}(t) \Struct{\qvec}(t + \tau) }.
  \label{equ:fourieramp}
\end{equation}
For $\hff = 0$, the 
optimal chain time $\tchain$ again corresponds to $\lchain \sim L$, and long 
chains ($\lchain \gg L$) again oscillate slowly (compare with \fig{fig:mode}). 
Without factor fields, long-wavelength excitations are much slower to 
decorrelate (see \fig{fig:densf}). The factor field again appears to lower the 
dynamical scaling exponent of density fluctuations, and by the same token, of 
overall correlations. 
  
\section{ECMC dynamics in the coexisting-phase regime}
\label{sec:FullDynamicsMixed}

At density $\eta=0.708$, the \twod hard-sphere system presents coexisting fluid 
and hexatic phases of different densities. In the two-phase region, the lower 
limit for the dynamical scaling exponent of a local MCMC algorithm is $z = 1$ 
because switching from one of the coexisting to the other in an extensive test 
volume requires mass transport of \bigObs{N} spheres by a distance \bigObs{L}. 
One of the coexisting phases, the hexatic, has quasi-long-range order, for which 
one expects $z \sim 2$ for Hamiltonian dynamics~\cite{Zippelius1980} and 
possibly also for ECMC. The hydrodynamic modes discussed in 
\sect{sec:FullDynamicsLiquids} are no longer the only slow ones. The correlation 
times of local MCMC algorithms have however not been firmly established in the 
coexisting-phase region and in the hexatic phase. Our preliminary computations 
in this section present evidence that in the coexisting-phase regime factor 
fields certainly do not greatly speed up the convergence. It is thus likely that 
the factor fields, as presently formulated, do not couple to the orientational 
order. 

\begin{figure}[!tbp]
  \includegraphics[width=0.9  \columnwidth]{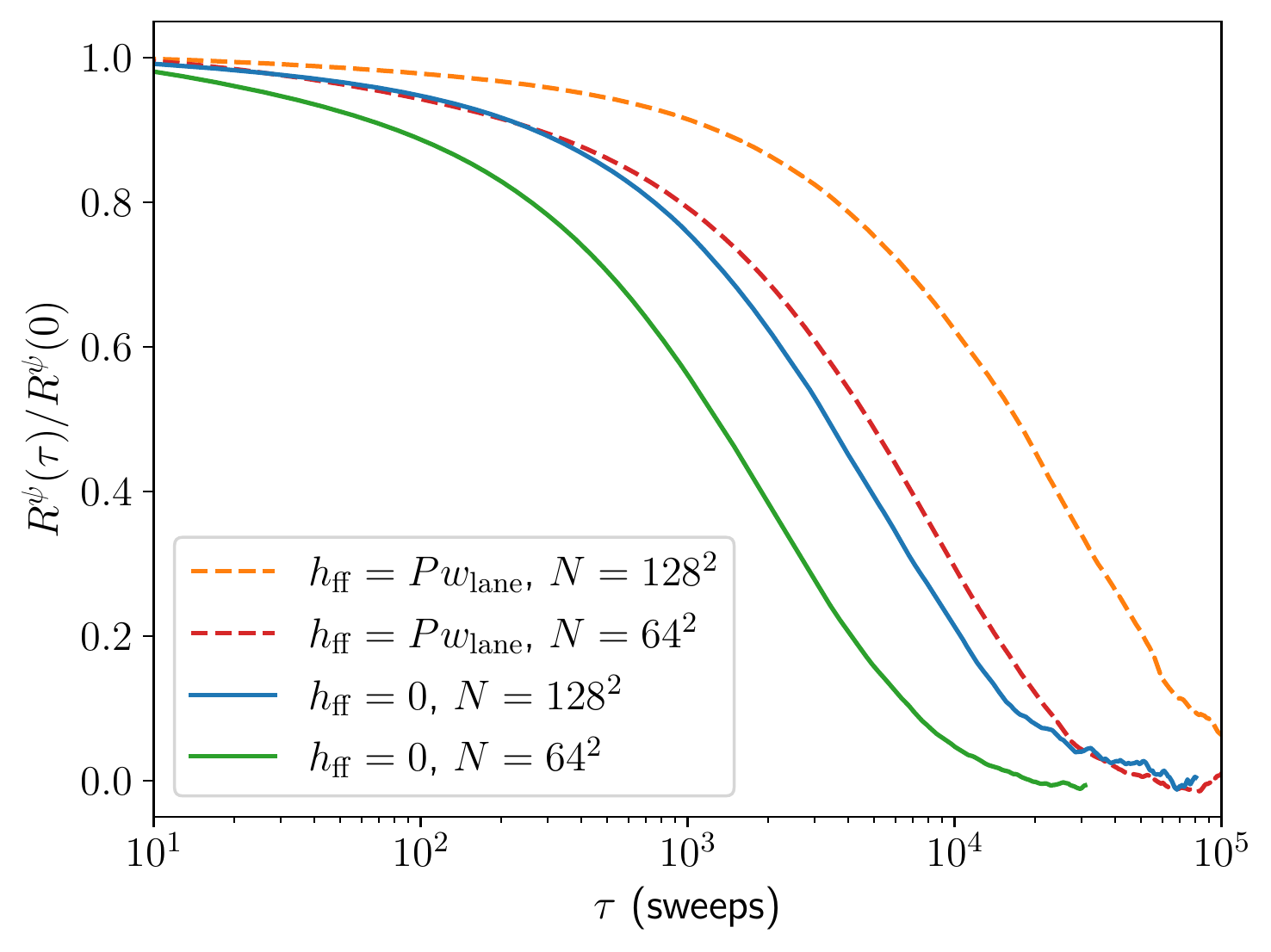}
\caption{ Autocorrelation  $R^{\psi}(\tau)$ of the squared amplitude $|\psi|^2$ 
of the global orientational order parameter in the two-phase region ($\eta = 
0.708$). The autocorrelations decay on a time scale that increases with $N$ both 
for $\hff=0$ and for $\hff= P \wlane$, where the simulation is slightly slowed 
down.}
\label{fig:psi}
\end{figure}

We evaluate the global orientational order parameter 
\begin{equation}
  \psi = \frac{1}{N}\sum_{j=1} ^{N}
\frac{1}{n(j)}\sum_{\text{neig:}p} e^{6 i 
\theta_{jp} }, 
\end{equation}
where $n(j)$ denotes the number of neighbors $p$ of \PARTICLE $j$, and 
$\theta_{jp}$ is the angle between \PARTICLES $j$ and $p$. Rather than the 
Voronoi classification  of neighbors we simply determine neighborhood through a 
cut-off based on the cell system. This does not change qualitative features.  We 
study the autocorrelation of the norm of this complex field, \fig{fig:psi}, 
which is sensitive to fluctuations in the amplitude of the hexatic field.
\begin{equation}
  R^{\psi}(\tau) = \mean{|\psi|^2(t) \,\,  |\psi|^2(t+ \tau)}
  \label{equ:psiamp}
\end{equation}

We perform simulation for $\hff= 0$ and for $\hff = P \wlane$ (see 
\fig{fig:psi}). Unlike the case of the density fluctuations in the fluid phase 
we find that the factor field slows the dynamics of the orientational order 
parameter $\psi$ by a small numerical factor. Even though long-wavelength 
density fluctuations are sampled efficiently by factor-field ECMC, this 
efficiency does not seem to feed into the dynamics of the global orientational 
order parameter.
  
\section{Conclusions}

In this paper, we have generalized factor-field ECMC from the previously 
introduced \oned case, where it appears natural, to higher-dimensional particle 
systems. As an example, we have implemented it for \twod hard spheres, although 
the new algorithm is trivial to extend to smooth interactions, as for example 
Lennard-Jones or soft-sphere potentials in any spatial dimension. Factor-field 
ECMC is found to sample density fluctuations very quickly, with a dynamical 
scaling exponent at the theoretical minimum $z = 0$ (in the \twod fluid phase), 
while reversible local Markov chains feature $z=2$ and molecular dynamics with coupling to a thermostat $z=1$. 
The autocorrelation times is reached once each sphere has moved a number of 
times that at most grows with the logarithm of the system size.

We have further discussed the factor-field algorithm at densities where \twod 
hard spheres present two coexisting phases of different densities, namely the 
fluid and the hexatic and demonstrated that local MCMC algorithms must have a 
dynamical scaling exponent $z \ge 1$, simply because the density differences 
require important  mass transfers between any two independent equilibrium 
configurations. Moreover, as one of the coexistent phases has quasi-long-range 
orientational order, we expect local algorithms to satisfy $z =2$. Simulation 
timescales involved in studying \twod hard spheres in the hexatic phase are 
still today extremely time-consuming, and the dynamical scaling exponents have 
not yet been computed. Our preliminary studies however do not allow us to 
conclude to any speed increases of factor-field ECMC in the presence of a 
hexatic phase. We conjecture that the factor field does not couple to
orientational degrees of freedom, because it is aligned with the chain 
direction. Modified non-reversible Markov chains that couple to hexatic order 
can be set up, possibly with factor fields that point in a direction 
different from the chain direction.

Besides its theoretical interest, we
imagine applications of factor-field ECMC (already in its present formulation) 
in the physics of glasses, as well as in studies of the melting transition for 
soft spheres, in \twod and higher. Most importantly, factor-field ECMC 
outperforms molecular dynamics, and it has superior dynamical scaling. This fact
was already proven in \oned (see \cite{Lei2019}) and is now firmly 
established in higher-dimensional systems. It should motivate further studies 
in non-reversible Markov chains. 

\bibliography{General,ff}
\end{document}